\documentclass[superscriptaddress,pre,twocolumn]{revtex4}
\usepackage{graphicx}
\usepackage{color}
\usepackage{amsmath,amsfonts,amssymb} 
\newcommand{\lastequal}{Corresponding authors. These authors contributed equally.}

\newcommand{\br}{\mathbf{r}}
\newcommand{\bW}{\mathbf{W}}
\newcommand{\beq}{\begin{equation}}
\newcommand{\eeq}{\end{equation}}
 
\begin{document}
\title{Physical observables to determine the nature of membrane-less cellular sub-compartments}

\author{Mathias Luidor Heltberg}
\affiliation{Laboratoire de physique de l'\'Ecole normale sup\'erieure,
  CNRS, PSL University, Sorbonne Universit\'e, and Universit\'e de
  Paris, 75005 Paris, France}
  \affiliation{Institut Curie,
  CNRS, PSL University, Sorbonne Universit\'e, 75005 Paris, France}
  \author{Judith Min\'e-Hattab}
  \affiliation{Institut Curie,
  CNRS, PSL University, Sorbonne Universit\'e, 75005 Paris, France}
    \author{Angela Taddei}
  \affiliation{Institut Curie,
  CNRS, PSL University, Sorbonne Universit\'e, 75005 Paris, France}
\author{Aleksandra M. Walczak}
\thanks{\lastequal}
\affiliation{Laboratoire de physique de l'\'Ecole normale sup\'erieure,
  CNRS, PSL University, Sorbonne Universit\'e, and Universit\'e de
  Paris, 75005 Paris, France}
\author{Thierry Mora}
\thanks{\lastequal}
\affiliation{Laboratoire de physique de l'\'Ecole normale sup\'erieure,
  CNRS, PSL University, Sorbonne Universit\'e, and Universit\'e de
  Paris, 75005 Paris, France}
\date{\today}

\begin{abstract}
The spatial organization of complex biochemical reactions is essential for the regulation of cellular processes. Membrane-less structures called foci containing high concentrations of specific proteins have been reported in a variety of contexts, but the mechanism of their formation is not fully understood. Several competing mechanisms exist that are difficult to distinguish empirically, including liquid-liquid phase separation, and the trapping of molecules by multiple binding sites. Here we propose a theoretical framework and outline observables to differentiate between these scenarios from single molecule tracking experiments. In the binding site model, we derive relations between the distribution of proteins, their diffusion properties, and their radial displacement. We predict that protein search times can be reduced for targets inside a liquid droplet, but not in an aggregate of slowly moving binding sites. 
These results are applicable to future experiments and suggest different biological roles for liquid droplet and binding site foci.

\end{abstract}

\maketitle
\section{Introduction}
The cell nucleus of eukaryotic cells is not an isotropic and homogeneous environment. In particular, it contains membrane-less sub-compartments, called foci or condensates, where the protein concentration is enhanced for certain proteins. Even though foci in the nucleus have been observed for a long time, the mechanisms of their formation, conservation and dissolution are still debated \cite{strom2017phase,altmeyer2015liquid,larson2017liquid,patel2015liquid,boehning2018rna,Pessina2019,McSwiggen2019a,McSwiggen2019,oshidari2020dna,Gitler2020,Erdel2020}. 
An important aspect of these sub-compartments is their ability to both form  at the correct time and place, and also to dissolve after a certain time. One example of foci are the structures formed at the site of a DNA double strand break (DSB) in order to localize vital proteins for the repair process  at the site of a DNA break \cite{lisby2001rad52}. Condensates have also been reported to be involved in gene regulation \cite{Hnisz2017,Bing2020} and in the grouping of telomeres in yeast cells \cite{Meister2013a,Ruault2021}.

\begin{figure*}
	\begin{center}
          \includegraphics[scale=0.3]{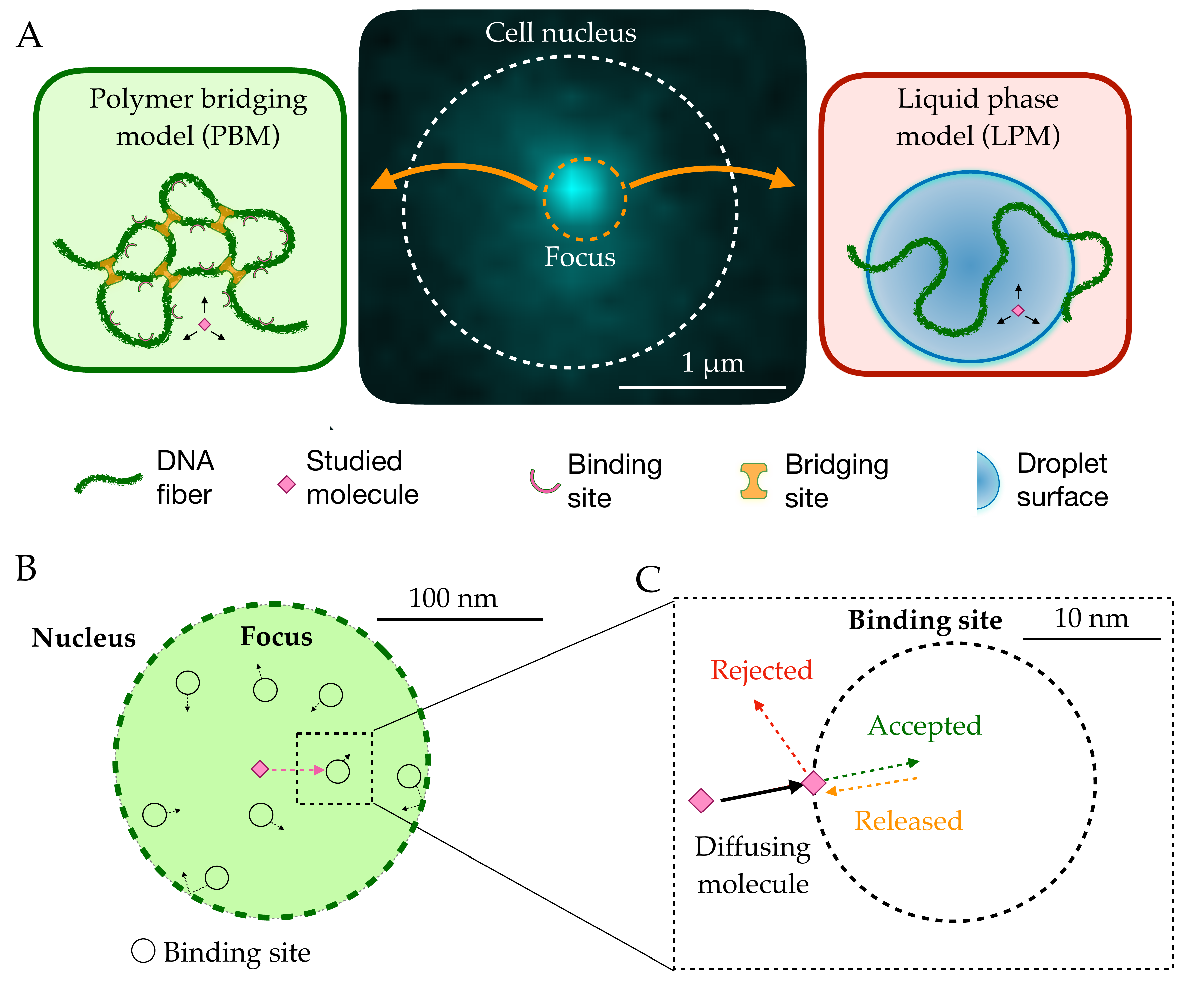}
	\caption{ 
	{\bf A.} In the middle, the observed signal from a fluorescently tagged  Rad52  protein inside the nucleus following a double stand break. Left: Schematic figure showing the Polymer Bridging Model (PBM). Proteins binding specifically to the chromatin stabilize it, effectively trapping the motion of other molecules. Right: Schematic figure showing the Liquid Phase Model (LPM). Liquid-liquid phase separation results in the formation of a droplet foci with a different potential and different effective diffusion properties than outside the droplet. {\bf B.} Details of the PBM model. Particles diffuse freely with diffusivity $D_n$ until they hit one of the $N$ spherical binding sites, themselves diffusing with diffusivity $D_b$. The focus is formed due a high concentration of binding sites. The binding sites are only partially absorbing, so that not all collision events result in a binding even. Once bound, the particle stays attached to the binding site, and then unbinds with rate $k_-$.
	\label{FIG1}
      }
      \end{center}
    \end{figure*}

Different hypotheses have been put forward to explain focus formation in the context of chromatin, among which two main ones (discussed in the particular context of DSB foci in \cite{Mine-Hattab2019}): the Polymer Bridging Model (PBM) and the Liquid Phase Model (LPM). The Polymer Bridging Model is based on the idea that specific proteins form bridges between different chromatin loci by creating loops or by stabilizing interactions between distant loci on the DNA (Fig.~\ref{FIG1}A, left). 
 These interactions can be driven by specific or multivalent weak interactions between chromatin binding proteins and chromatin components. In this case, the existence of sub-compartments relies on both the binding and bridging properties of these proteins. By contrast, the LPM posits that membrane-less sub-compartments arise from a liquid-liquid phase separation. In this picture, first proposed for P granules involved in germ cell formation~\cite{Brangwynne2009}, proteins self-organize into liquid-like spherical droplets that grow around the chromatin fiber, allowing certain molecules to become concentrated while excluding others (Fig.~\ref{FIG1}A, right).  
 
 Although some biochemical and wide field microscopy data support the LPM hypothesis for DSB foci \cite{altmeyer2015liquid, larson2017liquid, strom2017phase,McSwiggen2019a,Mine-Hattab2019}, these observations are at the optical resolution limit, and a more direct detection of these structures is still missing. Coarse-grained theoretical models of the LPM exist~\cite{statt2020model,grmela1997dynamics}, but predictions of microscale behaviour that can be combined with a statistical analysis of high resolution microscopy data to discriminate between the hypotheses has not yet been formulated. Previously, we analyzed in detail single-particle tracking data in the context of yeast DSB foci, and discussed their comptability with each model \cite{mine2021single}. Here, we build a general physical framework for understanding and predicting the behaviour of each model under different regimes. The framework is general and applicable to many different types of foci, although we chose to focus on the regime of parameters relevant to yeast DSB foci, for which we can directly related our results to experimental measurements.
While the LPM and PBM models have often been presented in the literature as opposing views, here we show under what conditions the PBM may be reduced to an effective description that is mathematically equivalent to the LPM, but with specific constraints linking its properties. We discuss the observables of the LPM and PBM and derive features that can be used to discriminate these two scenarios.

\section{Results}
\subsection*{Two models of foci}\label{Model}
To describe the situation measured in single particle tracking experiments, we consider the diffusive motion of a single molecule within the nucleus of a cell in the overdamped limit, described by the Langevin equation in 3 dimensions:
\beq\label{eq:langevin}
  d\br=dt\left[\nabla D(\br) -\frac{D(\br)}{k_BT}\nabla U(\br)\right] + \sqrt{2D(\br)}d\bW,
\eeq
where $\bW$ is a 3-dimensional Wiener process, $U(\br)$ is the potential exerted on the particle, and $D(\br)$ is a position-dependent diffusion coefficient. The $\nabla U$ term corresponds to a force divided by the drag coefficient $k_BT/D(\br)$, which is given in terms of $D$ and temperature according to Einstein's relation. The $\nabla D$ term comes from working within the It\^o convention. The steady state distribution of particles is given by Boltzmann's law:
\beq
p(\br)=\frac{1}{Z}\exp\left[-\frac{U(\br)}{k_BT}\right],
\eeq
where $Z$ is a normalization constant.

In the LPM, we associate the focus with a liquid  droplet characterized by a sudden change in the energy landscape.  We model the droplet as a  change in the potential $U(r)$, and a change in the diffusion coefficient $D(r)$ inside the droplet focus compared to the diffusion coefficient in the rest of the nucleus $D_n$. 
We assume both the diffusion coefficient and the potential are spherically symmetric around the center of the focus, and have sigmoidal forms:
\begin{align}
  D(r) & = D_0 + \frac{D_n-D_0}{1+e^{-b(r-r_f)}}, \label{sig1} \\
  U(r) & = \frac{A}{1+e^{-b(r-r_f)}},\label{sig2}
\end{align}
where $D_0$ is the diffusion coefficient inside the focus, $r_f$ is the radial distance to the center of the focus, and the coefficients are defined in Table~\ref{tab:parameters}. While this description is general, different relations between the diffusion coefficient and the surface potential are possible.

In the PBM we describe the dynamics of particles using a microscopic model (Fig.~\ref{FIG1}B and C). The focus has $N$ binding sites, each of which is a partially reflecting sphere \cite{bryan1891note,duffy2015green,carslaw1992conduction} (Fig. \ref{FIG1}C) with radius $r_b$. Binding sites can themselves diffuse with diffusion coefficient $D_b$, and are confined within the focus by a potential $U_b(\br)$, so that their density is $\rho(\br)\propto e^{-U_b(\br)}$ according to Boltzmann's law.
While not bound, particles diffuse freely with diffusion constant $D_n$, even when inside the focus. However, the movement of the particle is affected by direct interactions with the binding sites. Binding is modeled as follows. As the particle crosses the spherical boundary of a binding site during an infinitesimal time step $dt$, it gets absorbed with probability $p_b=\kappa \sqrt{{\pi\, dt}/{D_n}}$ (Fig.~\ref{FIG1}C), where $\kappa$ is an absorption parameter consistent with the Robin boundary condition at the surface of the spheres, $D\mathbf{n}\cdot\nabla p(\mathbf{x})=\kappa p(\mathbf{x})$ \cite{erban2007reactive,singer2008partially}, where $\mathbf{x}$ is a point on the surface of the sphere, and $\mathbf{n}$ is the unit vector normal to it.

While bound, particles follow the motion of their binding site, described by:
\beq
  d\br=-dt \frac{D_b}{k_BT}\nabla U_b(\br) + \sqrt{2D_b}d\bW,
\eeq
where $\bW$ is a 3-dimensional Wiener process. 
A bound particle is released with a constant rate $k_-$. If we exclude the binding site potential $U_b$, which is only relevant near the focus boundary to ensure consistency, the PBM has 5 parameters: $N$, $r_b$, $D_b$, $\kappa$ and $k_-$. Their typical values can be found in Table ~\ref{tab:parameters}. 

\begin{table*}
    \centering
    \begin{tabular}{c | c| l| c| c| c| c}
        Variable & Model & Description & Value & Range & Exp. value & Units \\ \hline\hline
        $r_f $ & both& radius of focus      & $100 $          & 50-200    & & ${\rm nm}$   \\
        $r_n $ &both& radius of nucleus      & $500$          & 300-1000 &   & ${\rm nm}$               \\
        $D_n$ & both& Diffusion coefficient in nucleus & 1.0 & 0.5-2.0 & 1.08 & ${\rm \mu m}^2/s$               \\
                $\sigma$ & both & Experimental noise level & 30 & 30 & 30 & $ {\rm nm} $ \\
        $D_0$ & LPM& Diffusion coefficient inside droplet & 0.05 & 0.01-0.5 &0.032 & ${\rm \mu m}^2/s$ \\
        $A$ & LPM & Surface potential & 5.0 & 0-10 & 5.5 & $k_BT$\\
        $b$ & LPM &Steepness in potential & 1000 & 500-10000 &  & ${\rm \mu m}^{-1}$ \\
        $\rho$   & PBM  & Density of binding sites inside focus     & $4.8\cdot 10^4$ & $1\cdot 10^3$-$8.4\cdot 10^4$ & & ${\rm \mu m}^{-3}$\\
        $D_b $ & PBM& Diffusion coefficient of binding sites      & $0.005 $        & $0$-$0.1$    & 0.005 & ${\rm \mu m}^2s^{-1} $  \\
         $r_b $ & PBM & Radius of binding sites      & $10 $          & $5$-$20$  &  & {\rm nm}            \\
         $k_{-} $ & PBM & Unbinding rate      & $500 $          & $10$-$10000$  &  & $s^{-1} $           \\         
      $\kappa $ & PBM & Absorption parameter      & $100 $          & $0$-$1000$ &   & ${\rm \mu m}/s$\\

    \end{tabular}
    \caption{Parameters used in this study with their typical values, and the ranges we have considered. Experimental values are from \cite{mine2021single}. $D_0$ and $A$ are model parameters in the LPM, but also effective observables in the PBM. The diffusivity of binding sites is taken to be that of Rfa1 molecules in the focus, which bind to single-stranded DNA in repair foci, and are thus believe to follow the diffusion of the chromatin \cite{mine2021single}.}
    \label{tab:parameters}
\end{table*}

\begin{figure*}
	\centering
	\includegraphics[scale=0.35]{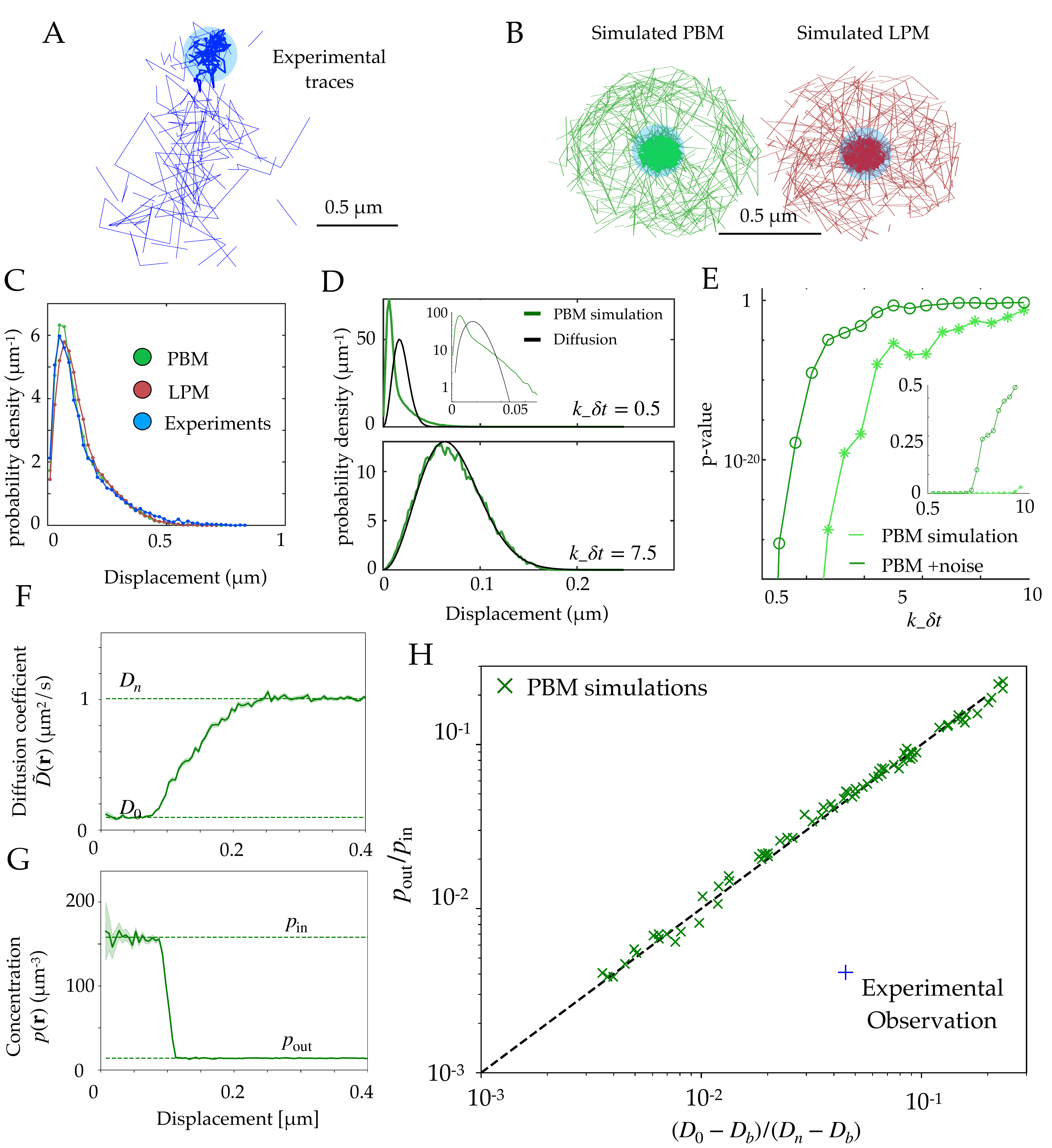}
	\caption{ 	
	{\bf A.} Example of experimental tracking of a Rad52 molecule visiting a double-strand break (DSB) locus.
	{\bf B.} Example trajectory of a particle visiting the focus from simulations in the PBM (left) and LPM (right). The simulated trajectories are visually similar to the data in B.
	{\bf C.} Displacement histogram (jump sizes) for the PBM, LPM and experiments, for an interval $\delta=20$ ms.
	{\bf D.} Displacement histogram for the PBM for small values of $k_- dt$ (top) and fast values (bottom). Here we varied the interval from $\delta t=1$ ms (bottom) to $\delta t=15$ ms (top).
	{\bf E.} Hypothesis testing using a two sided KS-test, comparing the displacement histogram of a free diffusion process (black line in D) and the displacement histogram of diffusion inside the focus (green line in D). Parameters are the same as in D. $\delta t$ was varied from $1$ to $25$ ms.
	{\bf F.} Effective diffusion coefficient as a function of distance to the focus center $r$, estimated from simulations of the PBM using displacement histograms.
	{\bf G.} Particle density $p(\br)$ as a function of $r$, estimated from simulations of the PBM.
	{\bf H.} Relation between the ratio $({D_0-D_b})/({D_n-D_b})$ versus the ratio of densities inside and outside the focus, both estimated from simulations of the PBM (green crosses), compared to the identity prediction (Eq.~\ref{eqn:Itopr}, black line). Blue cross shows the experimental observation for Rad52 in DSB loci from \cite{mine2021single}. Parameter values as in Table \ref{tab:parameters} except: $r_n=1\mu$m for B; $A=2.5k_BT$ for B-C; $r_n=0.3\ {\rm \mu m}$, $r_f=0.15\ {\rm \mu m}$ and $D_n=0.5\ {\rm\mu m}^2/s$ for D-E, $\kappa=300\ {\mu m}/s$ for D, $r_n=0.75\ {\rm \mu m}$ for F-H. In H we varied $\kappa=1$--$400\ {\mu m}/s$, $k_-=1$--$1,500$ s${}^{-1}$, and $\rho=2.4$--$4.8\cdot 10^4\ {\rm \mu m}$.
	}
	\label{FIG2}
\end{figure*}

\subsection*{Comparison between simulated and experimental traces}
In recent experimental work~\cite{mine2021single}, we used single particle tracking to follow the movement of Rad52 molecules, following a double-strand break in {\em S. cerevisiae} yeast cells, which causes the formation of a focus. These experiments show that temporal traces of Rad52 molecules concentrate inside the focus, as shown for a representative cell in Fig.~\ref{FIG2}A.

Using both the PBM and LPM models described above,  we can construct traces that look similar to the data (Fig.~\ref{FIG2}B). To mimick the data, we only record and show traces in two dimensions and added detection noise corresponding to the level reported in the experiments~\cite{mine2021single}. Based on these simulations, we gather the statistics of the particle motion to create a displacement histogram representing the probability distribution of the observed step sizes between two successive measurements. For this choice of parameters, both of the models and the experimental data  look very similar (Fig.~\ref{FIG2}C).

In principle, we could have expected the displacement histogram of particles inside the focus to look markedly different between the PBM and the LPM. While the LPM should follow the prediction from classical diffusion (given by a Gaussian radial distribution, $p(|\delta \br|)\propto |\delta \br|^2 e^{-|\delta \br|^2/(4D\delta t)}$ for an interval $\delta t$), the PBM prediction is expected to be in general non-Gaussian  because of intervals during which the particle is bound and almost immobile (as the chromatin or single-stranded DNA carrying the binding sites moves very slowly), creating a peak of very small displacements.
Simulations show that departure from Gaussian displacements is most pronounced when the binding and unbinding rates are slow compared to the interval $\delta t$ (Fig.~\ref{FIG2}D, top), but is almost undetectable when they are fast (Fig.~\ref{FIG2}D, bottom). With our parameters, the binding rate is $k_+\rho \approx 3,000$ s${}^{-1}$, and $k_-$ ranges from $10$ to $10,000$ s${}^{-1}$, with $\delta t=20$ ms. For comparison, assuming weak binding to DNA, $K_d=k_+/k_-\approx 1\ {\rm \mu M}$ would give $k_-\sim 40$ s${}^{-1}$, and assuming strong specific binding, $K_d\sim 1$ nM, implies $k_-\sim 0.04$ s${}^{-1}$.
Fig.~\ref{FIG2}E shows how the detectability of non-Gaussian displacements gets worse as $k_-\delta t$ increases, and is further degraded by the presence of measurement noise.

The experimental findings of single Rad52 molecules in yeast repair foci \cite{mine2021single} suggest that the movement inside the focus are consistent with normal diffusion (Fig.~\ref{FIG2}C). While this observation excludes a wide range of slow binding and unbinding rates in the PBM, it does not rule out the PBM itself. In addition, separating displacements inside the focus from boundary-crossing ones can be very difficult in practice, and errors in that classification may result in spurious non-Gaussian displacement distributions that would confound this test. Therefore, it is important to find observables that can distinguish the two underlying models.

\subsection*{Effective description of the Polymer Binding Model}
Motivated by experimental observations, we analyze the PBM in a mean-field description, which is valid in the limit where binding and unbinding events are fast relative to the traveling time of the particles. 
In this regime, a particle rapidly finds binding sites with rate $k_+\rho(\br)$ (where $\rho(\br)$ is the density of binding sites) and unbinds from them with rate $k_-$. While in principle rebinding events complicate this picture, they can be renormalized into a lower effective unbinding rate \cite{kaizu2014berg}. Assuming that interactions between binding sites do not affect their binding to the particle of interest, the binding rate can be approximated in the presence of partially reflecting binding sites by the Smoluchowski rate \cite{nadler1996reaction,berezhkovskii2019trapping} (Appendix A):
\beq
k_{+}=\frac{4\pi D_n r_b}{1+\frac{D_n}{r_b\kappa}}.
\eeq
Since the processes of diffusion, binding, and unbinding are in equilibrium, the steady state distribution of a particle can be derived using Boltzmann's law. At each position $\br$, the unbound state is assigned weight 1, and the bound state weight $\rho(\br)/K_d$, where $K_d=k_-/k_+$ is the dissociation constant. Then the probability distribution of the particle's position is given by:
\beq\label{eq:pr}
p(\br)\propto\left(1+\frac{\rho(\br)}{K_d}\right)\propto \frac{1}{p_{u}(\br)},
\eeq
where
\beq
p_{u}(\br)=\frac{k_-}{k_-+k_+\rho(\br)}
\eeq
is the probability of being unbound conditioned on being at position $\br$.

In the limit of fast binding and unbinding, the dynamics of particles are governed by an effective diffusion coefficient, which is a weighted average between the free diffusion of tracked molecules, and the diffusion coefficient of the binding sites:
\begin{align}
\tilde D(\br) = p_u (\br) D_n + (1-p_u (\br)) D_b= \frac{D_n k_{-} + D_bk_{+}\rho(\br)}{k_{-} +k_{+} \rho(\br)}.\label{EqDr}
\end{align}
Likewise, particles are pushed by an effective confinement force: when they are bound to binding sites, they follow their motion which is confined inside of the focus. The resulting drift is given by that of the binding sites, but weighted by the probability of being bound to them:
\beq
\begin{split}
\langle d\br\rangle&=-dt(1-p_u(\br))\frac{D_b}{k_BT}\nabla U_b(\br)\\
&=dt\left[-\frac{\tilde D(\br)}{k_BT}\nabla \tilde U(\br)+\nabla \tilde D(\br)\right],
\end{split}
\eeq
where in the second line we have rewritten the dynamics in terms of an effective potential $\tilde U(\br)=k_BT\ln(1+k_+\rho(\br)/k_-)$, using $\rho(\br)\propto e^{-U_b(\br)/k_BT}$.
Thus the effective dynamics may be described by the Langevin equation of the same form as the LPM (\ref{eq:langevin}) but with the relation between $\tilde U(\br)$ and $\tilde D(\br)$ constrained by their dependence on $\rho(\br)$:
\beq\label{UvsD}
\tilde U(\br) = k_BT \ln\left[\frac{\tilde D(\br)-D_b}{D_n-D_b}\right],
\eeq
with the convention that $\tilde U=0$ far away from the focus where $\rho=0$.
As a consistency check, one can verify that the equilibrium distribution $p\propto e^{-\tilde U/k_BT}$ gives back Eq.~\ref{eq:pr}.

\subsection*{Scaling relation between concentration and diffusivity in the PBM}

Experiments or simulations give us access to the {\it effective diffusivity} through $\tilde{D} = \langle \delta \br^2\rangle/ (2d\delta t)$, where $\delta t$ is the time between successive measurements, and $d$ the dimension in which motion is observed.
Within the PBM, Eq.~\ref{UvsD} allows us to establish a general relation between the particle concentration $p(r)$, which can also be measured, and the effective diffusivity $\tilde{D}$, through:
\beq
\label{eqn:Itopr}
p(\br) \propto \frac{1}{\tilde{D}(\br) - D_b}.
\eeq
Typically in experiments we have $D_b\ll \tilde D\ll D_n$, in which case this relation may be approximated by $p(\br)\tilde D(\br)={\rm const}$.

We validated Eq.~\ref{eqn:Itopr} in simulations of the PBM. We divided the radial coordinate $r$ into small windows of $10^{-3} {\rm \mu m}$ and plotted the measured effective diffusion coefficient $\tilde{D}(r)$, as a function of $r$ (Fig.~\ref{FIG2}F), as well as the density of tracked particles $p(r)$ (Fig.~\ref{FIG2}G). The diffusivity increases from $D_0$ inside the focus to $D_n$ outside, while the density decreases from $p_{\rm in}$ to $p_{\rm out}$. We extracted those values numerically from the simulations.
Fig.~\ref{FIG2}H shows that Eq.~\ref{eqn:Itopr} predicts well the relationship between these 4 numbers, for a wide range of parameter choices of the PBM (varying $\kappa$ from 1 to 400 ${\rm \mu m}/s$, $k_{-}$ from 5 to 1500 $s^{-1}$ and $\rho$ from $23873-47746{\rm \mu m_{-3}}$, while keeping $D_b = 5\cdot 10^{-3} {\rm \mu m}^2/s$ and the other parameters to values given by Table~\ref{tab:parameters}). While this relation was derived in the limit of fast binding and unbinding, it still holds for the slower rates explored in our parameter range. However, it breaks down in the limit of strong binding, when we expect to see two populations (bound and unbound), making the effective diffusion coefficient an irrelevant quantity.

We can compare this prediction to estimates from the experimental tracking of single Rad52 molecules in yeast repair foci \cite{mine2021single}, assuming that the diffusivity of the binding sites is well approximated by that of the single-stranded DNA-bound molecule Rfa1, measured to be $D_b = 5\cdot 10^{-3} {\rm \mu m}^2/s$. This experimental point, shown as a blue cross in Fig.~\ref{FIG2}H, substantially deviates from the PBM prediction: Rad52 particles spend much more time inside the focus than would be predicted from their diffusion coefficient based on the PBM.
To agree with the data, the diffusion coefficient of binding sites would have to be increased to $D_b=0.0314 {\rm \mu m}^2/s$, which is almost an order of magnitude larger than what was found in experiments.

\begin{figure*}
	\centering
	\includegraphics[width=\linewidth]{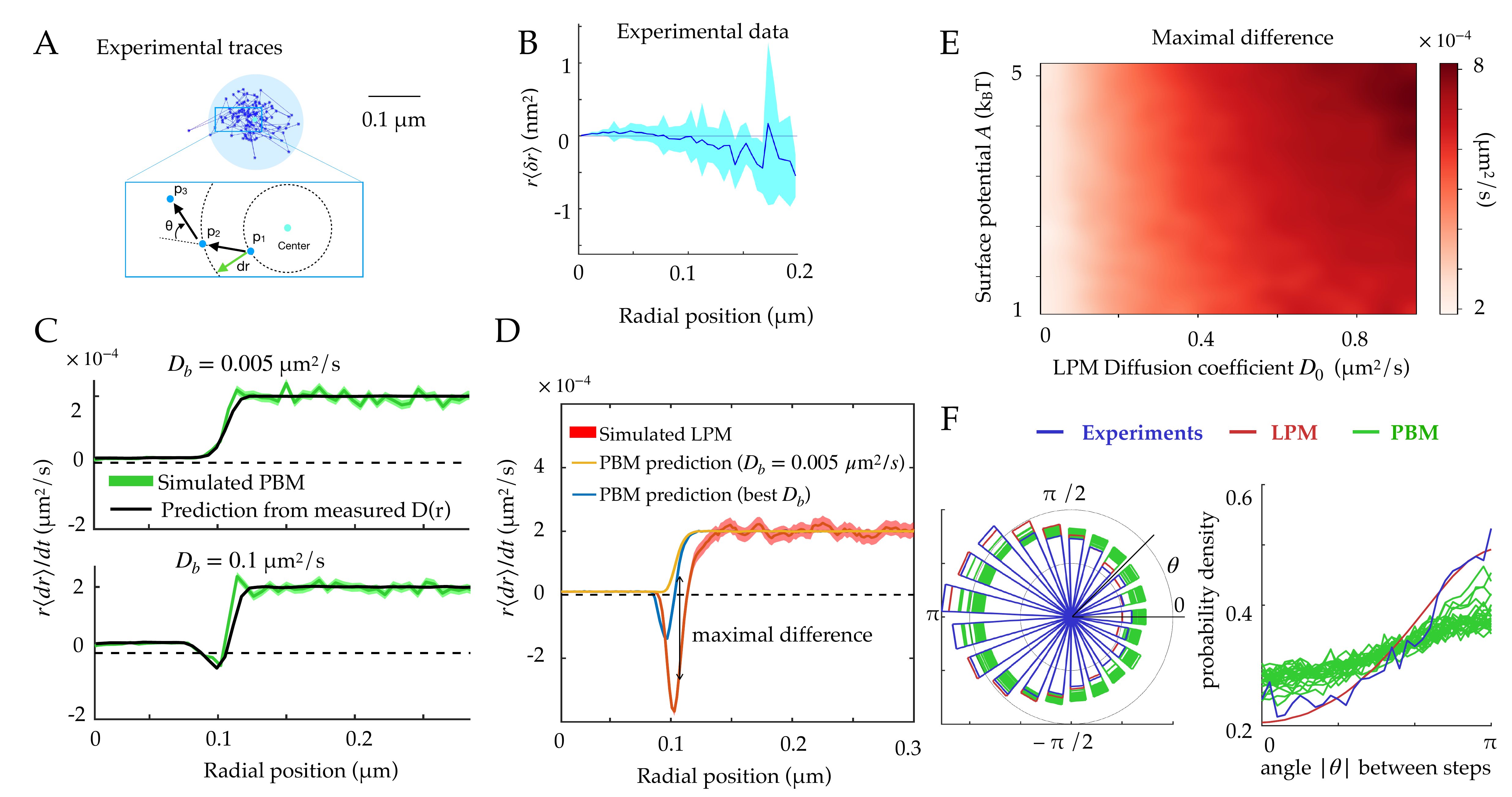}
	\caption{ 
	{\bf A.} Experimental traces of Rad52 in a DSB locus~\cite{mine2021single}, showing traces close to the focus boundary. The inset shows the definition of the radial movement $\delta r$, where a particle moves a specific distance away from the center of the focus, as well as the angle $\theta$ between consecutive displacements.
	{\bf B.} Data extracted from experiments~\cite{mine2021single} to estimate the average radial displacement of the tracked particle multiplied by the radius. 
	{\bf C.} Simulations showing the radial displacements in the PBM with slowly (top) and rapidly (bottom) moving binding sites. Black lines are predictions based on the measurement of $\tilde D$ (Eq.~\ref{eqn:Itopr}). 
	{\bf D.} Radial displacememnt from simulations of the LPM. Black line shows the (wrong) prediction made while assuming the PBM, using the measurement of the effective diffusion coefficient $\tilde D(r)$. We call the discrepancy between data and the PBM prediction the ``Maximal positive difference.''
	{\bf E.} Heatmap showing the maximal positive difference in LPM simulations as a function of $D_0$ and $A$.
 {\bf F.} Distribution of angles (represented radially on the left, and linearly on the right) between the displacements of consecutive steps of length $\delta t$, from experiments and simulations. Multiple curves for the PBM correspond to different parameter choices corresponding to the points of Fig.~2H.
        Parameter values as in Table \ref{tab:parameters}, except $r_n=1\ \mu$m for F. For F parameters are varied with $\kappa=100$--$300\ \mu$m/s, $k_-=500$--$1,500$ s${}^{-1}$, $r_f=0.1$--$0.14$ $\mu$m.
	}
	\label{FIG3}
\end{figure*}

\subsection*{Diffusion coefficient and concentration predict boundary movement in the PBM}
Another observable that is accessible through simulations and experiments is the radial displacement near the focus boundary.
In practice, we gather experimental traces around the focus, and estimate the radius of the focus as shown in Fig.~\ref{FIG3}A. Using many traces, we can find the average radial displacement $\langle\delta r\rangle$ during $\delta t$, as a function of the radial position of the particle $r$ (Fig.~\ref{FIG3}B). Under the assumption of spherical symmetry, within the PBM this displacement is given by:
\beq\label{eqn:Itodr}
\begin{split}
  \langle dr\rangle &=  dt \left(-(1-p_u(r))\frac{D_b}{k_BT}\partial_r U_b(r) + \frac{\tilde D(r)}{r}\right)\\
  &=dt\left(-\frac{\tilde D(r)}{k_BT}\partial_r \tilde U(r)+\partial_r \tilde D(r) +\frac{\tilde D(r)}{r}\right),
  \end{split}
\eeq
where the term $\tilde D/r$ comes from the change to spherical coordinates.

The first line of Eq.~\ref{eqn:Itodr} shows that the average change in radial position of single particles $\langle dr\rangle$ cannot be negative in the PBM for steady binding sites ($D_b=0$).
This is reproduced in simulations, for different absorption probabilities, as shown in Fig.~\ref{FIG3}C.

By constrast, in the LPM there is no constraint on the sign of the displacement $\langle dr\rangle$ since the relation between the diffusion coefficient and the 
surface potential is not constrained like in the PBM.
Even when binding sites can move, this prediction can be used to falsify the PBM.
Eq.~\ref{eqn:Itodr} makes a prediction for the average radial displacement of the tracked molecule in the PBM, solely as a function of the diffusivity and concentration profiles $\tilde D(r)$ and $p(r)$, using $\tilde U(r)=k_B T\ln p(r)$. Accordingly, this prediction agrees well with simulations of the PBM (Fig.~\ref{FIG3}C).

Using Eq.~\ref{eqn:Itodr} that is derived for the PBM, along with the definition of $\tilde U$ as a function of $\tilde D$ in Eq.~\ref{UvsD}, to analyze a simulation of the LPM leads to large disagreement between the inferred and true parameters. This PBM-based analysis underestimates the depth of the potential
(Fig.~\ref{FIG3}D). It predicts a negative displacement $\langle dr\rangle$ when $D_b$ is inferred using the PBM formula $p_{\rm in}/p_{\rm out}=(D_n-D_b)/(D_f-D_b)$, although its magnitude is underestimated. But when taking the experimental value of $D_b=0.005\ {\rm \mu m}^2/s$, $\langle dr\rangle$ is always positive even at the boundary.
This spurious entropic ``reflection'' is an artifact of using the wrong model, and is distinct from the Laplace pressure, which only affects macroscopic objects.
The inference using the PBM of such a positive displacement at the surface of the focus can therefore be used to reject the PBM.
Fig.~\ref{FIG3}E represents the magnitude of that discrepancy as a function of two LPM parameters ---\,diffusivity inside the droplet and surface potential\,--- showing that the PBM is easier to reject when diffusivity inside the focus is high.

In summary, the average radial diffusion coefficient can predict the radial displacement of tracked molecules within the PBM, and deviations from that prediction can be used as a means to reject the PBM using single-particle tracking experiments.

\subsection*{Distribution of angles between consecutive time steps}

To go beyond the average radial displacement, we considered a commonly used observable to study diffusive motion in complex environment: the distribution of angles between two consecutive displacements during $\delta t$. While this distribution is uniform in 2 dimensions \cite{Liao2012}, it is expected to be asymmetric in presence of confinement and obstacles \cite{Izeddin2014}.

We computed this distribution from simulations of the PBM and LPM, and compared them to experiments in yeast repair foci (Fig.~\ref{FIG3}F). These distributions are all asymmetric, with an enrichment of motion reversals (180 degree angles). Since the LPM assumes standard diffusion within a potential, the asymmetry in that model can be entirely explained by the effect of confinement, which tends to push back particles at the focus boundary.
With the parameters of Table \ref{tab:parameters}, the LPM agrees best with the data, while the PBM shows a more moderate asymmetry across a wide range of parameters. This is perhaps counter-intuitive: because of the interaction of tracked molecules with the binding sites, which can hinder and reflect their motion, one expects the PBM, but not the LPM, to display angle asymmetry even in the absence of a confinement boundary. This expectation is confirmed by simulating the PBM in an infinite focus with a constant density of binding sites (Fig.~S1). However, this asymmetry is seen only  when the measurable time step is small or comparable to the binding time, and must be further corrected for the effect of confinement, which makes it unfit to discriminate between the two models in the context of yeast repair foci.

\begin{figure*}
	\centering
	\includegraphics[scale=0.45]{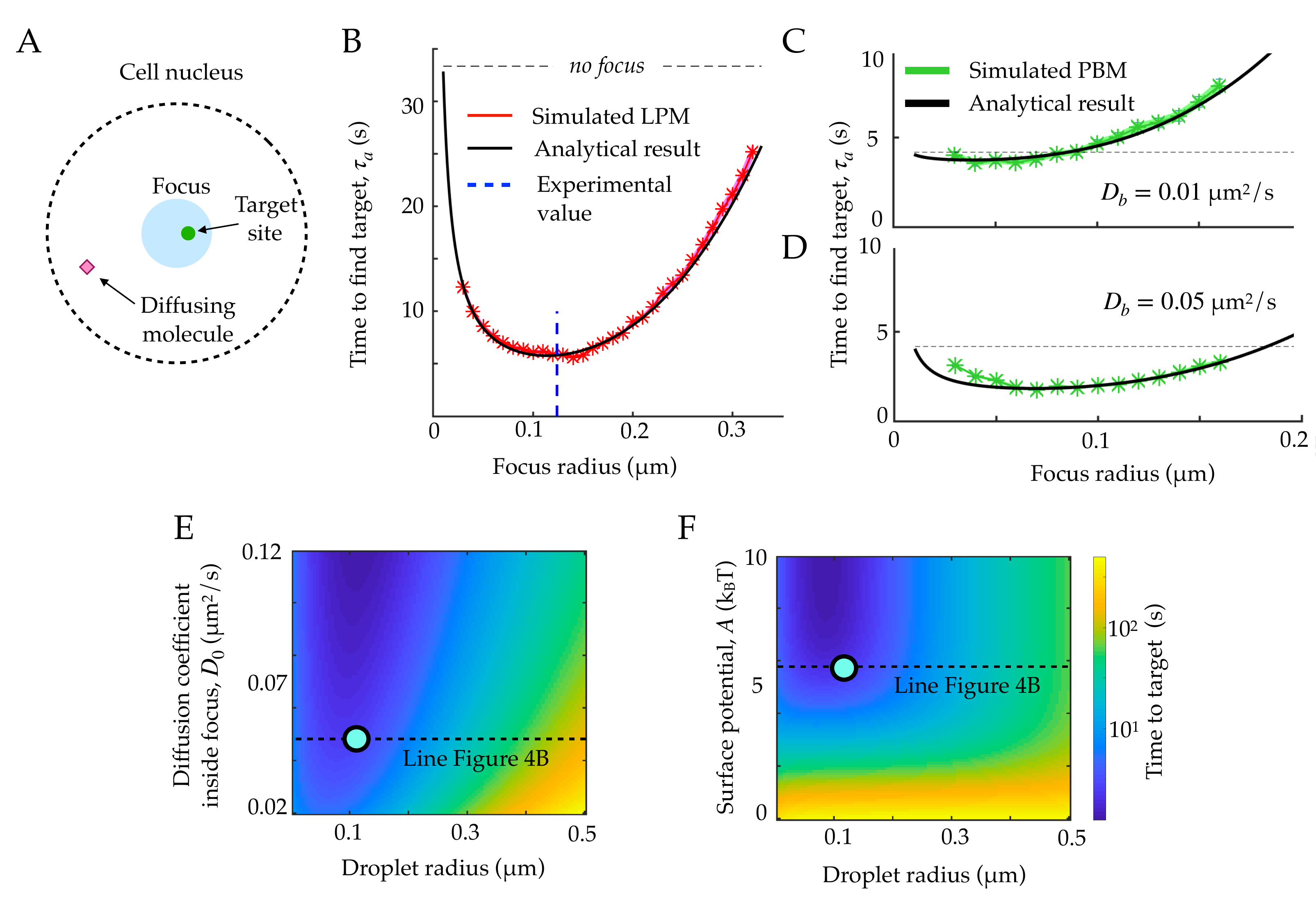}
	\caption{
	{\bf A.} Schematic figure showing the setup of the tracked molecule and the effective target. 
	{\bf B.} Time to reach the specific target in simulations for the LPM. Black curve shows the predicted result from the analytical derivation. Here we use parameters: $A=-5.5k_BT$, $D_0 = 0.05 \mu^2/s$, $r_n = 1.0 {\rm \mu m}$, $D_n = 0.8 \mu^2/s$.
	{\bf C.} Same as B, but for the PBM. Same parameters as in B, but with $D_0 = D_n\cdot e^{U(r)/k_BT}$.
	{\bf D.} Heatmap showing the expected search time as a function of the droplet size (x-axis) and the focus diffusion coefficient (y-axis). Green point corresponds to experimental observations.
	{\bf E.} Heatmap showing the expected search time as a function of the droplet size (x-axis) and the height of the surface potential (y-axis). Green point corresponds to experimental observations.
        Parameter values as in Table \ref{tab:parameters}, except $b=2,000\ {\rm \mu m}^{-1}$, $D_0=0.04\ {\rm \mu m}^2/s$, $A=5.5k_BT$ for B, E and {\bf F}, $\kappa=50\ {\rm \mu m}^2/s$, $k_-=100$ s${}^{-1}$, $\rho=2.4\cdot 10^4\ {\rm \mu m}^{-3}$ for C-D.
	}
	\label{FIG4}
\end{figure*}

\subsection*{Foci accelerate the time to find a target, but only moderately in the PBM}
Foci keep a higher concentration of molecules of interest within them through an effective potential.
We wondered if this enhanced concentration of molecules could act as a ``funnel'' allowing molecules to find their target (promoter for a transcription factor, repair site, etc) faster.

To address this question, we consider an idealized setting with spherical symmetry, in which the target is a small sphere of radius $r_0$ located at the center of the focus, of radius $r_f$ (Fig.~\ref{FIG4}A). We further assume that the nucleus is a larger sphere of radius $r_n$, centered at the same position. We start from a general Langevin equation of the form in Eq.~\ref{eq:langevin}, and assume that the target is perfectly absorbing, creating a probability flux $J=\tau_{a}^{-1}$, equal to the rate of finding the target for a single particle. The corresponding Fokker-Planck equation can be solved at steady state, giving (Appendix B):
\beq
\tau_a=\int_{r_0}^{r_n}dr\, r^2e^{-U(r)/k_B T}\int_{r_0}^r \frac{dr'}{D(r')r'^2}e^{U(r')/k_BT}.
\eeq
Taking the particular form of Eqs.~\ref{sig1} and \ref{sig2}, with a sharp boundary $br_f\gg 1$, the integral can be computed explicitly:
\beq
\begin{split}
\tau_a=& \frac{r_f^3-r_0^3}{3D_0r_0} + \frac{r_0^2 - r_f^2}{2D_0} + e^{-\frac{A}{k_BT}}\Big(\frac{r_n^3-r_f^3}{3D_0r_0} + \frac{r_f^3 - r_n^3}{3D_0r_f} \Big) \\
&+ \frac{r_n^3-r_f^3}{3D_nr_f} +\frac{r_f^2-r_n^2}{2D_n}. \label{Full_Time}
\end{split}
\eeq

In the limit $r_0\ll r_f\ll r_n$ and of a strong potential $A\gg k_BT$, Eq.~\ref{Full_Time} simplifies to:
\beq
\tau_a\approx \frac{r_f^3}{3D_0r_0} +\frac{r_n^3}{3D_nr_f},\label{tau_a_approx}
\eeq
which is exactly the sum of the time it takes to find the focus from the edge of the nucleus, and the time it takes to find the target from the focus boundary. 

Expression \eqref{Full_Time} can be related to the celebrated Berg and Purcell bound \cite{berg1977physics}, which sets the limit on the accuracy of sensing small ligand concentration by a small target, due to the limited number of binding events during some time $t$. With a mean concentration of ligands $c$ in the cell nucleus, there are $m=(4\pi/3) r_n^3c$ such ligands, and their rate of arrival at the target is $m/\tau_a={4\pi cr_n^3}/{(3\tau_a)}$, so that the number of binding events during $t$ is equal to $n\sim {4\pi cr_n^3t}/{(3\tau_a)}$ on average. Random Poisson fluctuations of $n$ result in an irreducible error in the estimate of the concentration $c$:
\beq
\frac{\delta c^2}{c^2}\sim\frac{\delta n^2}{n^2}\sim \frac{1}{n}\sim \frac{3\tau_a}{{4\pi cr_n^3t}}.\label{BP_first}
\eeq
Replacing $\tau_a$ in Eq.~\ref{BP_first} with the expression in Eq.~\ref{tau_a_approx}, we obtain in the limit of large nuclei ($r_n\to \infty$):
\beq
\frac{\delta c^2}{c^2}\sim \frac{1}{4\pi ct}\left[\frac{1}{D_nr_f}+\frac{e^{-A/k_BT}}{D_0}\left(\frac{1}{r_0}-\frac{1}{r_f}\right)\right].
\eeq
One can further check that in the limit of a strong potential, or when there is no focus, $r_0=r_f$, we recover the usual Berg and Purcell limit for a perfectly absorbing spherical measurement device, $\delta c/c\sim 1/\sqrt{4\pi D_ncr_ft}$.

Eq.~\ref{Full_Time} agrees well with simulations in the general case (Fig.~\ref{FIG4}B),  where we used parameters obtained for Rad52 in a repair focus~\cite{mine2021single}. 
Eq.~\ref{Full_Time} typically admits a minimum as a function of $r_f$, meaning that there exists an optimal focus size that minimizes the search time.
Using the measured parameters for Rad52, we find an optimal focus size of $r_f^* \approx 120$ nm, which matches the estimated droplet size $r_f =124$ nm in  these experiments \cite{mine2021single}  (dashed line in Fig.~\ref{FIG4}B).
In the limit where $r_n\gg r_0$, the optimal size takes the explicit form:
\beq
(r_f^*)^4=r_0r_n^3 \frac{\frac{D_0}{D_n}-e^{-\frac{A}{k_BT}}}{3(1-e^{-\frac{A}{k_BT}})}=r_0r_n^3 \frac{\frac{D_0}{D_n}-\frac{p_{\rm out}}{p_{\rm in}}}{3\left(1-\frac{p_{\rm out}}{p_{\rm in}}\right)}.
\eeq
This optimum only exists for $D_0e^{A/k_BT}>D_n$ or $D_0p_{\rm in}>D_np_{\rm out}$, that is, when the benefit of spending more time in the focus compensates the decreased diffusion coefficient. Incidentally, in that case the Berg and Purcell bound on sensing accuracy generalizes to:
\beq
\frac{\delta c^2}{c^2}\sim \frac{1}{\pi ct}\left(\frac{p_{\rm out}}{4p_{\rm in} D_0r_0}+\frac{1}{3D_nr_f}\right).
\eeq

The previous formulas for the search time and sensing accuracy are valid for the general Langevin equation (\ref{eq:langevin}), which describes both the LPM and the PBM in the mean-field regime. Fig.~\ref{FIG4}C and D show the search time as a function the focus size for the specific case of the PBM, where diffusion and potential are further linked.
The relation between $\tilde U$ and $\tilde D$, given by Eq.~\ref{UvsD}, imposes
$D_0e^{A/k_BT}=D_n + D_b(e^{A/k_BT}-1)>D_n$, giving the optimal focus size:
\beq
(r_f^*)^4=r_0r_n^3 \frac{D_b}{3D_n}.
\eeq
For the physiologically relevant regime of very slow binding sites, $D_b\ll D_n$, this optimal focus size shrinks to 0, meaning that the focus offers no benefit in terms of search time, because binding sites ``sequester'' or ``titre out'' the molecule, preventing it from reaching its true target.

These results suggest to use the search time, or equivalently the rate for binding to a specific target, as another measure to discriminate between the LPM and the PBM. Since only the LPM gives a clear minimum of the search time as a function of focus size, identifying an optimal focus size would rule out the PBM. Conversely, a monotonic relation between the search time and the focus size would be consistent with the PBM (without excluding the LPM). Testing for the existence of such a minimum would require experiments where the focus size may vary, and where reaching the target can be related to a measurable quantity, such as gene expression onset in the context of gene regulation.

\section{Discussion}

The PBM and the LPM are the two leading physical models for describing the nature of nuclear foci or sub-compartments.
In this work, we analyzed how the traces of single particle tracking experiments should behave in both models. 
Using statistical mechanics, we derived a mean field description of the PBM that shares the general functional form of the LPM (Eq.~\ref{eq:langevin}), but with an additional constraint linking concentration and diffusion inside the focus: the denser the focus, the higher the viscosity. This constraint does not appear to be satisfied by the experimental data on Rad52 in repair foci, favoring the liquid droplet hypothesis.
We use our formulation of the PBM to predict the behaviour of the mean radial movement around the focus boundary, which may differ markedly from observation of traces inside a liquid droplet (described by the LPM).
We find the range of LPM parameters where this difference would be so significant that it would lead to ruling out the PBM. This work provides a framework for distinguishing the LPM and PBM, and should be combined with modern inference techniques to accurately account for experimental noise and limited data availability (for instance accounting for molecules going out of the optimal focus). Future improvements in single-particle tracking experiments will allow for longer and more accurate traces necessary to deploy the full potential of these methods.

The LPM and PBM have often been presented as opposing models~\cite{Mine-Hattab2019}, driven by attempts to compare the macroscopic properties of different membraneless sub-compartments to the original example of liquid-like P granules~\cite{Brangwynne2009}. The LPM is a macroscopic description of a liquid droplet in the cytoplasm \cite{Hyman2014}, which condensates molecules inside the droplet, and alters their different diffusion properties. The droplet is formed by a phase transition, which means it will be recreated if destroyed, and will go back to its spherical shape if sheared or merged. Conversely, the PBM is a microscopic description that provides an explicit bridging mechanism by which a focus is formed. Here we clarified the link between the two from the point of view of single molecules. We confirmed mathematically the intuition that, in the limit of very fast binding and unbinding, the PBM is a particular case of the LPM model. Going further, we show that the PBM prescription imposes a strong constraint between the effective diffusion of molecules in the sub-compartment, $D(\br)$, and the effective potential, $\tilde U(\br)$ (Eq.~\ref{UvsD}). The LPM is compatible with this choice, but does not impose it in general, although alternative mechanistic implementations of the LPM may impose similar constraints with different functional forms. The correspondance between the two models breaks down when binding and unbinding are slow. However, for this regime to be relevant, experimental observations need to be fast enough to capture individual binding or unbinding events, which is expected to be hard in general, and was not observed in the case of repair foci in yeast.

We found another way in which the two models behave very differently: in the LPM, the focus may act as a funnel accelerating the search for a target inside the focus, and we calculated the optimal focus size that minimizes the search time. In the PBM, such an improvement is negligible unless binding sites themselves have a fast diffusive motion. This difference between the two models could potentially be tested in experiments where the focus size varies. It is not clear whether this optimality argument is relevant for DSB: the merger of two foci leads to larger condensates, suggesting that the focus size is not tightly controlled.
But the argument may be relevant for gene expression foci, especially in the context of development where transcription factors need to reach their regulatory target fast in order to ensure rapide cell-fate decision making \cite{Bialek2019}.
On the contrary, if a focus is created in order to decrease the probability of specific binding, such as in silencing foci \cite{Brown1997}, a PBM implementation may be more advantageous.
Binding sites, which act as decoys~\cite{Burger2009}, sequester proteins involved in gene activation, thus increasing the time its takes them to reach their target and suppressing gene expression. In that picture, genes would be regulated by the mobility and condensation of these decoy binding sites.

More generally, foci or membranelss sub-compartments are formed in the cells for very different reasons and remain stable for different timescales. For example, repair foci are formed for short periods of time (hours) to repair double strand breaks, and then dissolve. In this case the speeds of both focus formation and target finding are important for rapid repair, but long term stability of foci is not needed. Gene expression foci~\cite{Hnisz2017,Bing2020} can be long lived, and their formation may be viewed as a way to ``prime'' genes for faster activation. However, given the high concentrations of certain activators, not all genes may require very fast search times of the transcription factors to the promoter. While molecularly the same basic elements are available for foci formation -- binding and diffusion -- different parameter regimes exploited in the LPM and PBM may lead to different behaviour covering a vast range of distinct biological requirements.

\section{Methods}
\subsubsection*{Simulation of PBM}
In order to simulate the bridging model we generated $N$ binding sites of radius $r_b$.
We simulate a diffusing molecule through the free overdamped Langevin equation in 3 dimensions, and at each time-step we 
find the closest binding site to the particle. If the distance of the particle ($\Delta r$) is smaller than $r_b$ we bind the molecule with probability 
$p_{b} = \kappa \sqrt{{\pi \delta t}/{D_n}}$.
If the particle does not bind, it is reflected so the new distance to the center of the particular binding site is $2r_b - \Delta r$.
At this new position we evaluate the position of all other binding sites (they all diffuse with diffusion coefficient $D_b$, and if the molecule is within the radius of another binding site (happens extremely rarely), it is again accepted to bind with the same probability $p_b$.
If a particle binds, it stays at the position of the intersection with the binding site, and at each time step it can be released with probability $k_-\delta t$. We choose $\delta t$ small so that $p_b \ll 1$ and $\sqrt{2 D_n \delta t} \ll r_b$, which for the considered parameter ranges in Table~1 is typically obtained for values  of $\delta t = 10^{-6} s$.

\subsubsection*{Simulation of LPM}
To simulate the LPM, we use the Milstein algorithm to calculate the motion of a particle. As in the PBM, the particle is reflected at the nucleus boundary, and can otherwise move freely in the nucleus. 
We typically choose the same value of $\delta t$ as the PBM, since the surface potential typically has a very steep gradient, given by $b\approx 1000$ as shown in Table I.

\subsection*{Acknowledgements}
The authors are grateful to Jean-Baptiste Masson, Alexander Serov and Ned Wingreen for valuable discussions. The study was supported by the Agence Nationale de la Recherche (Q-life ANR-17-CONV-0005), Centre National de la Recherche Scientifique (80' MITI project PhONeS), the European Research Council COG 724208,  the Labex DEEP (ANR-11-LABEX-0044 DEEP and ANR-10-IDEX-0001?02 PSL),  the ANR DNA-Life (ANR-15-CE12-0007),  the Fondation pour la Recherche M\'edicale (DEP20151234398), and the ANR-12-PDOC- 0035?01. The authors greatly acknowledge the PICT-IBiSA@Pasteur Imaging Facility of the Institut Curie, member of the France Bioimaging National Infrastructure (ANR-10-INBS-04).

\bibliographystyle{pnas}

\begin{thebibliography}{10}

\bibitem{strom2017phase}
Strom AR, {et~al.}
\newblock (2017) Phase separation drives heterochromatin domain formation.
\newblock \emph{Nature} 547:241--245.

\bibitem{altmeyer2015liquid}
Altmeyer M, {et~al.}
\newblock (2015) Liquid demixing of intrinsically disordered proteins is seeded
  by poly (adp-ribose).
\newblock \emph{Nature communications} 6:1--12.

\bibitem{larson2017liquid}
Larson AG, {et~al.}
\newblock (2017) Liquid droplet formation by hp1$\alpha$ suggests a role for
  phase separation in heterochromatin.
\newblock \emph{Nature} 547:236--240.

\bibitem{patel2015liquid}
Patel A, {et~al.}
\newblock (2015) A liquid-to-solid phase transition of the als protein fus
  accelerated by disease mutation.
\newblock \emph{Cell} 162:1066--1077.

\bibitem{boehning2018rna}
Boehning M, {et~al.}
\newblock (2018) Rna polymerase ii clustering through carboxy-terminal domain
  phase separation.
\newblock \emph{Nature structural \& molecular biology} 25:833--840.

\bibitem{Pessina2019}
Pessina F, {et~al.}
\newblock (2019) {Functional transcription promoters at DNA double-strand
  breaks mediate RNA-driven phase separation of damage-response factors}.
\newblock \emph{Nature Cell Biology} 21:1286--1299.

\bibitem{McSwiggen2019a}
McSwiggen DT, Mir M, Darzacq X, Tjian R
\newblock (2019) {Evaluating phase separation in live cells: diagnosis,
  caveats, and functional consequences}.
\newblock \emph{Genes {\&} development} 33:1619--1634.

\bibitem{McSwiggen2019}
McSwiggen DT, {et~al.}
\newblock (2019) {Evidence for DNA-mediated nuclear compartmentalization
  distinct from phase separation}.
\newblock \emph{eLife} 8:1--31.

\bibitem{oshidari2020dna}
Oshidari R, {et~al.}
\newblock (2020) Dna repair by rad52 liquid droplets.
\newblock \emph{Nature communications} 11:1--8.

\bibitem{Gitler2020}
Gitler AD, Shorter J, Ha T, Myong S
\newblock (2020) {Just Took a DNA Test, Turns Out 100{\%} Not That Phase}.
\newblock \emph{Molecular Cell} 78:193--194.

\bibitem{Erdel2020}
Erdel F, {et~al.}
\newblock (2020) {Mouse Heterochromatin Adopts Digital Compaction States
  without Showing Hallmarks of HP1-Driven Liquid-Liquid Phase Separation}.
\newblock \emph{Molecular Cell} 78:236--249.e7.

\bibitem{lisby2001rad52}
Lisby M, Rothstein R, Mortensen UH
\newblock (2001) Rad52 forms dna repair and recombination centers during s
  phase.
\newblock \emph{Proceedings of the National Academy of Sciences} 98:8276--8282.

\bibitem{Hnisz2017}
Hnisz D, Shrinivas K, Young RA, Chakraborty AK, Sharp PA
\newblock (2017) {A Phase Separation Model for Transcriptional Control}.
\newblock \emph{Cell} 169:13--23.

\bibitem{Bing2020}
Bing XY, Batut PJ, Levo M, Levine M, Raimundo J
\newblock (2020) {SnapShot: The Regulatory Genome}.
\newblock \emph{Cell} 182:1674--1674.e1.

\bibitem{Meister2013a}
Meister P, Taddei A
\newblock (2013) {Building silent compartments at the nuclear periphery: A
  recurrent theme}.
\newblock \emph{Current Opinion in Genetics and Development} 23:96--103.

\bibitem{Ruault2021}
Ruault M, {et~al.}
\newblock (2021) {Sir3 mediates long-range chromosome interactions in budding
  yeast}.
\newblock \emph{Genome research} 31:411--425.

\bibitem{Mine-Hattab2019}
Min{\'{e}}-Hattab J, Taddei A
\newblock (2019) {Physical principles and functional consequences of nuclear
  compartmentalization in budding yeast}.

\bibitem{Brangwynne2009}
Brangwynne CP, {et~al.}
\newblock (2009) {Germline P granules are liquid droplets that localize by
  controlled dissolution/condensation}.
\newblock \emph{Science} 324:1729--1732.

\bibitem{statt2020model}
Statt A, Casademunt H, Brangwynne CP, Panagiotopoulos AZ
\newblock (2020) Model for disordered proteins with strongly sequence-dependent
  liquid phase behavior.
\newblock \emph{The Journal of chemical physics} 152:075101.

\bibitem{grmela1997dynamics}
Grmela M, {\"O}ttinger HC
\newblock (1997) Dynamics and thermodynamics of complex fluids. i. development
  of a general formalism.
\newblock \emph{Physical Review E} 56:6620.

\bibitem{mine2021single}
Min{\'e}-Hattab J, {et~al.}
\newblock (2021) Single molecule microscopy reveals key physical features of
  repair foci in living cells.
\newblock \emph{Elife} 10:e60577.

\bibitem{bryan1891note}
Bryan G
\newblock (1891) \emph{Note on a problem in the linear conduction of heat}
\newblock Vol.{}~7, pp 246--248.

\bibitem{duffy2015green}
Duffy DG
\newblock (2015) \emph{Green's functions with applications}
\newblock (cRc press).

\bibitem{carslaw1992conduction}
Carslaw HS, Jaeger JC
\newblock (1992) \emph{Conduction of heat in solids}
\newblock (Clarendon press) No.{} BOOK.

\bibitem{erban2007reactive}
Erban R, Chapman SJ
\newblock (2007) Reactive boundary conditions for stochastic simulations of
  reaction--diffusion processes.
\newblock \emph{Physical Biology} 4:16.

\bibitem{singer2008partially}
Singer A, Schuss Z, Osipov A, Holcman D
\newblock (2008) Partially reflected diffusion.
\newblock \emph{SIAM Journal on Applied Mathematics} 68:844--868.

\bibitem{kaizu2014berg}
Kaizu K, {et~al.}
\newblock (2014) The berg-purcell limit revisited.
\newblock \emph{Biophysical journal} 106:976--985.

\bibitem{nadler1996reaction}
Nadler W, Stein D
\newblock (1996) Reaction--diffusion description of biological transport
  processes in general dimension.
\newblock \emph{The Journal of chemical physics} 104:1918--1936.

\bibitem{berezhkovskii2019trapping}
Berezhkovskii AM, Dagdug L, Bezrukov SM
\newblock (2019) Trapping of diffusing particles by small absorbers localized
  in a spherical region.
\newblock \emph{The Journal of chemical physics} 150:064107.

\bibitem{Liao2012}
Liao Y, Yang SK, Koh K, Matzger AJ, Biteen JS
\newblock (2012) {Heterogeneous single-molecule diffusion in one-, two-, and
  three-dimensional microporous coordination polymers: Directional, trapped,
  and immobile guests}.
\newblock \emph{Nano Letters} 12:3080--3085.

\bibitem{Izeddin2014}
Izeddin I, {et~al.}
\newblock (2014) {Single-molecule tracking in live cells reveals distinct
  target-search strategies of transcription factors in the nucleus}.
\newblock \emph{eLife} 2014:1--27.

\bibitem{berg1977physics}
Berg HC, Purcell EM
\newblock (1977) Physics of chemoreception.
\newblock \emph{Biophysical journal} 20:193--219.

\bibitem{Hyman2014}
Hyman AA, Weber CA, J{\"{u}}licher F
\newblock (2014) {Liquid-Liquid Phase Separation in Biology}.
\newblock \emph{Annual Review of Cell and Developmental Biology} 30:39--58.

\bibitem{Bialek2019}
Bialek W, Gregor T, Tka{\v{c}}ik G
\newblock (2019) {Action at a distance in transcriptional regulation}.
\newblock \emph{arXiv:1912.08579}.

\bibitem{Brown1997}
Brown KE, {et~al.}
\newblock (1997) {Association of transcriptionally silent genes with Ikaros
  complexes at centromeric heterochromatin}.
\newblock \emph{Cell} 91:845--854.

\bibitem{Burger2009}
Burger A, Walczak AM, Wolynes PG
\newblock (2010) {Abduction and asylum in the lives of transcription factors}.
\newblock \emph{Proceedings of the National Academy of Sciences of the United
  States of America} 107:4016--4021.

\end{thebibliography}

\appendix

\onecolumngrid

\renewcommand{\thefigure}{S\arabic{figure}}
\setcounter{figure}{0}

\section{Binding rate by a partially absorbing sphere}
We consider a particle with diffusivity $D$, which can be partially absorbed by
a spherical binding site of radius $r_b$ and absorption parameter
$\kappa$. Its Fokker-Planck equation takes the following form, in spherical
coordinates projected onto the distance to the center of the binding
site, $r$:
\begin{align}\label{laplace}
\partial_t p = \frac{D}{r^2}\partial_r r^2\partial_r p.
\end{align}
The boundary conditions are $p(r=\infty)=1/V$, where $V$ is the total
volume, assumed to be much larger than that of the binding site, and
the Robin condition:
\beq
D\partial_r p(r_b)=\kappa p(r_b).
\eeq
The solution of Eq.~\ref{laplace} at steady state with these bondary
conditions reads:
\beq
p(r)=\frac{1}{V} \left(1-\frac{\kappa r_b/r}{\kappa+D/r_b}\right),
\eeq
the total diffusive flux is then given by
\beq
J=4\pi D r_b^2\partial_r p(r_b)= \frac{1}{V}\frac{4\pi D
  r_b}{1+\frac{D}{r_b\kappa}}.
\eeq
Normalizing by the volume factor gives the association rate for binding, $k_+=V J= 4\pi D
r_b/(1+D/\kappa r_b)$.

\section{Searching for a target in a funneling potential}
We consider a problem similar to that of the previous appendix. Now
the spherical object is a target, which is perfectly absorbing. It is
at the center of a liquid dropblet, which we model by a spherically symmetric potential $U(\br)$.

The probability distribution of a molecule is denoted by
$p(\br)=p(r)$. The probability density of being at distance $r$ from
the center, $q(r)$, is related to $p(r)$ through $q(r)=4\pi r^2 p(r)$,
accounting for the volume of the sphere.
The evolution of $r$ is described by the stochastic differential equation:
\beq
dr = \frac{2D}{r}+\partial_r D -\frac{D}{k_BT}\partial_r U +
\sqrt{2D}dW,
\eeq
where $W$ is a 1-dimensional Wiener process.
The corresponding Fokker-Planck equation reads:
\beq
\partial_t q = -\partial_r\left[\left(\frac{2D}{r}+\partial_r
  D-\frac{D}{k_BT}\partial_r
  U\right)q\right]+\partial_r^2(Dq)\doteq -\partial_r J.
\eeq
At steady state with a non-vanishing flux $J={\rm const}$, we have:
\beq
\left( \frac{2D}{r} - \frac{D}{k_BT}\partial_r U \right)q = D\partial_r q - J,
\eeq
or equivalently:
\beq
q\partial_r \phi + \partial_rq = \frac{J}{D}
\eeq
with $\phi\doteq -2\ln(r) + U/{k_BT}$. Multiplying both sides of the equation by $e^\phi$, we obtain:
\beq
\partial_r (e^\phi q) = \frac{J}{D}e^\phi.
\eeq
The general solution to that equation is:
\begin{align}
q(r) = Ce^{-\phi(r)} + Je^{-\phi(r)}\int_{r_0}^r
  \frac{e^{\phi(r')}}{D(r')}dr'.
\end{align}
We have $C=0$ because of the absorbing boundary condition $q(r_0)=0$. The
constant $J$ is determined by the normalization $\int_{r_0}^{r_n} dr\,
q(r)=1$, yielding:
\beq
J^{-1}\doteq \tau_a = \int_{r_0}^{r_n} dr\, e^{-\phi(r)}\int_{r_0}^r
dr'\, \frac{e^{\phi(r')}}{D(r')},
\eeq
This in turns gives the result of the main text after replacing
$\phi(r)$ by its definition.

\begin{figure*}
	\centering
	\includegraphics[scale=0.3]{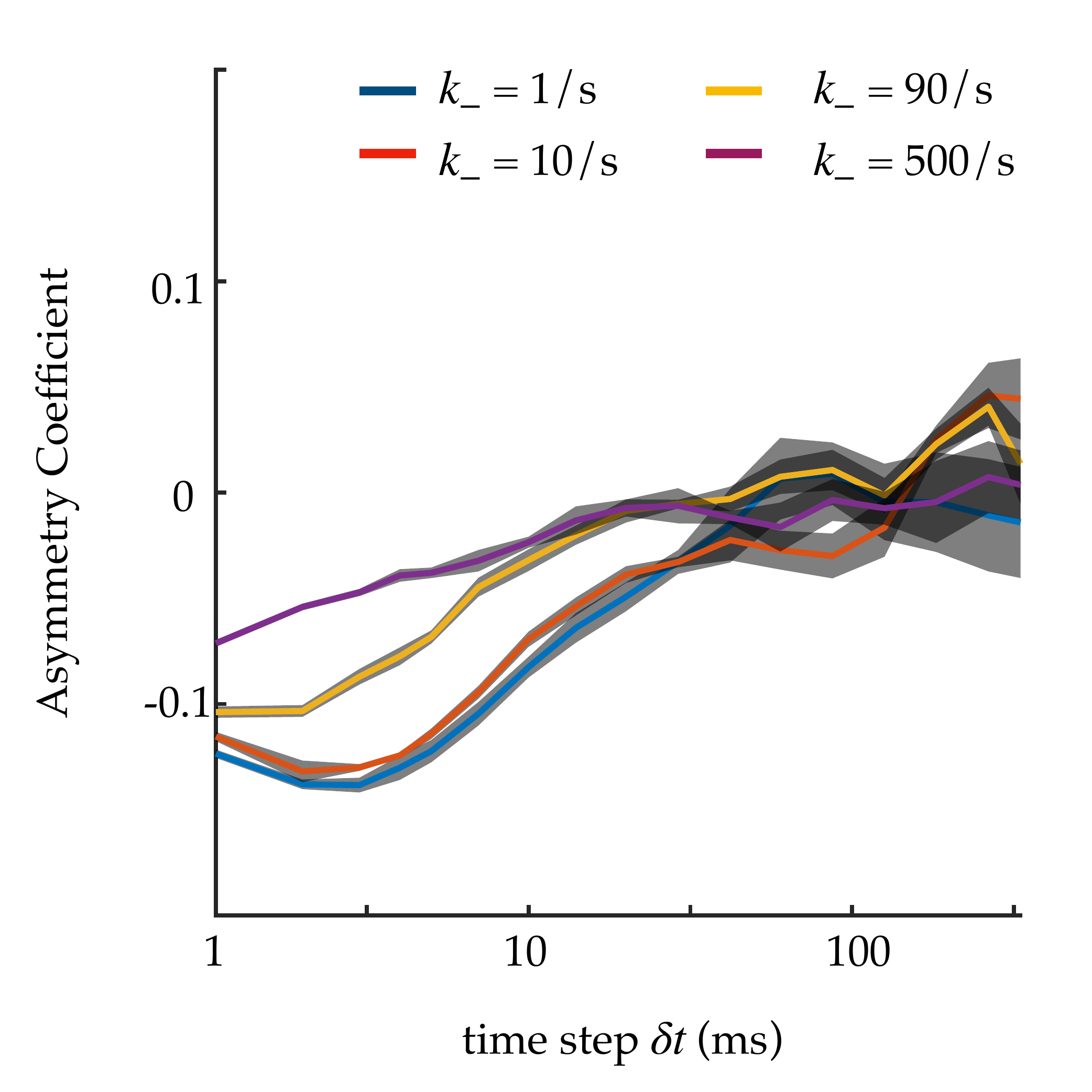}
	\caption{Asymmetry coefficient for an infinite focus, simulated
          with the PBM, as a function of the time step $\delta
          t$. Asymmetry is due to molecules reflecting off binding
          sites, causing more consecutive displacements to have 180
          degree angles. The faster the binding and unbinding relative
          to the time step $\delta t$, the closer to mean-field limit
          of standard diffusion, and the more symmetric the angle distribution.
          $k_+$ is changed alongside $k_-$ to keep $p_u$ constant. The
          asymmetry coefficient is defined as
          $\log_2[\mathbb{P}(|\theta|<\pi/6)/
          \mathbb{P}(|\theta|>5\pi/6)]$, where $-\pi<\theta<\pi$.
          Standard deviation is obtained from 4 independent runs of 1,000 s.
	}
	\label{FIGS1}
      \end{figure*}

\end{document}